\documentclass[aps,twocolumn,showpacs,prb,superscriptaddress,amsfonts,amsmath,amssymb]{revtex4-1}

\usepackage{graphicx}

\begin{document}

\title{T-matrix formulation of real-space dynamical mean-field theory \\ and the Friedel sum rule for correlated lattice fermions}

\author{K. Byczuk}
\email{byczuk@fuw.edu.pl}
\affiliation{Institute of Theoretical Physics,
Faculty of Physics,  University of Warsaw, ul.~Pasteura 5, PL-02-093 Warszawa, Poland}

\author{B. Chatterjee}
\affiliation{Institute of Physics, Czech Academy of Sciences, Na Slovance 2, 182 21 Prague, Czech Republic}

\author{D. Vollhardt}
\affiliation{Theoretical Physics III, Center for Electronic Correlations and Magnetism,
Institute of Physics, University of Augsburg, D-86135, Augsburg, Germany}

\date{\today}

\begin{abstract}
We formulate real-space dynamical mean-field theory within scattering theory. Thereby the Friedel sum rule is derived for interacting lattice fermions at zero temperature.

\end{abstract}


\maketitle

\section{Introduction}

In a metal the long-range Coulomb potential of a charge $Q$  is screened by the surrounding conduction electrons. For free electrons,  taking
into account that scattering is restricted to states at the Fermi surface, the screened potential and the electron density exhibit oscillations in position space. These ``Friedel oscillations'' \cite{Friedel52,Friedel58,Kittel} around a charged impurity decay algebraically with distance.
Charge neutrality requires that the number of electrons $Z_{\rm sc}$ participating in the screening of the electrostatic potential is equal to the charge of the impurity, i.e., $e Z_{\rm sc}+Q=0$, where $e$ is the electronic charge. The number $Z_{\rm sc}$ can be positive or negative and denotes the difference between the number of electrons with and without the impurity, respectively, i.e., it corresponds to the difference in the valence of the impurity and the host metal.
For non-interacting electrons, which experience the impurity only as an additional potential, Friedel proved a theorem which relates the screening charge $eZ_{\rm sc}$ to the scattering phase shifts in the  electron gas\cite{Friedel52,Friedel58} as
\begin{equation}
Z_{\rm sc}=\frac{2}{\pi}\sum_{l=0}^{\infty} (2l+1) \phi_l(\epsilon_F).
\label{Friedel_sum_rule}
\end{equation}
Here $\phi_l(\epsilon)$ is the scattering phase shift in the angular momentum channel $l$ at the Fermi energy $\epsilon_F$.\cite{Kittel} The Friedel sum rule (\ref{Friedel_sum_rule}) allows one to determine the scattering phase shifts of electrons at the Fermi energy. These phase shifts can be used to reconstruct both the asymptotic form of the wave function and the impurity potential. In fact, it is often sufficient to know only a few scattering phases rather than the full impurity potential itself.
The Friedel sum rule was later reformulated for lattice systems,\cite{Rudnick73,Mahan95} where the angular momentum is not a conserved quantum number, and also for interacting electrons,\cite{Langer61,Martin82} as well as for the single-impurity Anderson model.\cite{Langreth66} Nowadays Friedel oscillations and the Friedel sum rule play a significant role in many branches of condensed matter physics.\cite{Villain16,Georges16}
Separate Friedel sum rules hold for the total excess charge, spin and orbital momentum of the impurity in generic models for impurities in a metallic host. \cite{Yoshimori82} This concept is particularly relevant for transition metal impurities where the interaction between $d$ electrons and the $s-d$ hybridization is described by the Anderson model.
Furthermore, one can combine the Friedel sum rule with the Landauer formula to obtain a simple mathematical expression for the resistivity in symmetric molecular conductors which can be measured experimentally.\cite{Datta97}
It can also be used to understand the transport and thermodynamic properties in mesoscopic samples connected to leads, such as quantum interference effects in a single channel quantum wire in the presence of a point defect.\cite{Bandopadhyay03}
 Such studies may be useful for building future mesoscopic devices.
 The Friedel sum rule also finds its application in electronic interferometry to explain the dephasing in a detector at zero temperature due to the quantum fluctuations.\cite{Rosenow12}
In particular, the detection and characterization of  Friedel oscillations around inhomogeneities such as impurities, external interface  potentials, etc., in strongly correlated electron systems\cite{Sessi15} and fermionic cold atoms in optical lattices\cite{Riechers17} provide a deeper understanding of quasiparticle states in these systems.\cite{Dalla16}

A general method for the investigation of correlated lattice fermions is the dynamical mean-field theory (DMFT) \cite{Metzner89,Georges96}.
The DMFT is a comprehensive, non-perturbative, and diagrammatically controlled approximation scheme which allows one to study correlation phenomena even at intermediate coupling strengths, such as the Mott-Hubbard metal-insulator transition.
%
It was originally formulated for systems with discrete translational invariance. Consequently the DMFT self-consistency equations are expressed in Fourier (momentum) space. To take into account the effect of  inhomogeneities which break this invariance, DMFT was later reformulated in such a way that the self-consistency equations are completely expressed in real (lattice) space.\cite{Dobrosavljevic97,Potthoff99,Freericks04,Okamoto04,Helmes08,Snook08} This new formulation of DMFT is referred to as  "real-space dynamical mean-field theory"  (R-DMFT).

In this paper we develop a scattering formalism for correlated lattice fermions within R-DMFT.\cite{Dobrosavljevic97,Potthoff99,Freericks04,Okamoto04,Helmes08,Snook08}  This approach provides a natural setting for analytical and numerical treatments of inhomogeneous correlated lattice systems. As an application we generalize the Friedel sum rule for correlated particles in terms of many-body states at zero temperature. The Friedel sum rule obtained thereby is exact within R-DMFT and can serve as a test for approximations or numerical treatments of the R-DMFT equations.

\section{Inhomogeneous Hubbard model}

We consider the inhomogeneous Hubbard model
\begin{equation}
\hat{H}=\sum_{ij\sigma} t_{ij} \hat{a}^{\dagger}_{i\sigma} \hat{a}_{j\sigma} + U\sum_i \hat{n}_{i\uparrow} \hat{n}_{j\downarrow} + \sum_{i \sigma} V_{i\sigma} \hat{n}_{i \sigma},
\label{hamiltonian}
\end{equation}
where $\hat{a}^{\dagger}_{i\sigma}$ and $\hat{a}_{i\sigma}$ are creation and annihilation operators for fermions with spin $\sigma$ on  lattice site $i$, $\hat{n}_{i\sigma}=  \hat{a}^{\dagger}_{i\sigma} \hat{a}_{i\sigma}$  is the particle number operator at this site,  $t_{ij}$ is the hopping matrix element between the sites $i$ and $j$ with $t_{ii}=0$, $U$ is the local Hubbard  interaction between fermions of opposite spins on the same site $i$, and  the last term corresponds to an inhomogeneous external potential $V_{i\sigma}$ which is assumed to be local and real.

All one-particle properties as well as the thermodynamics of this model are determined by the one-particle Green function
\begin{equation}
G_{ij \sigma} (\tau) = - \langle T_{\tau} a_{i\sigma} (\tau) a^{\dagger}_{j\sigma}(0) \rangle,
\end{equation}
where $\tau$ is the imaginary  time  and the Heisenberg picture is used.  The average is taken within the grand canonical ensemble with the temperature $T=1/\beta$ and the chemical potential $\mu$. The Fourier transform yields the Green function $G_{\sigma}(i\omega_n)_{ij}$ with Matsubara frequency $\omega_n=(2n+1)\pi/\beta$.
In the following we only consider systems without long-range order.

\section{Real-space dynamical mean-field theory}

The model (\ref{hamiltonian}) will now be solved within R-DMFT.\cite{Dobrosavljevic97,Potthoff99,Freericks04,Okamoto04,Helmes08,Snook08}
For each lattice site $i$ the cavity method\cite{Georges96} gives the partition function
\begin{equation}
Z_i=Z^{(i)} \int D[a_{i\sigma}, a^*_{i\sigma}]e^{-S_i[a_{i\sigma}, a^*_{i\sigma}]},
\end{equation}
with the local action
\begin{align}
S_i[a_{i\sigma}, a^*_{i\sigma}]=-\int_0^{\beta}d \tau \int_0^{\beta}d \tau' a^*_{i\sigma} (\tau){\cal G}_{i\sigma}(\tau-\tau')^{-1} a_{i\sigma} (\tau) \nonumber \\
 + U \int_0^{\beta}d \tau \; n_{i\uparrow} (\tau)n_{j\downarrow}(\tau) ,
\end{align}
where $Z^{(i)}$ is the partition function of the system without site $i$.
The Weiss Green function ${\cal G}_{i\sigma}(i\omega_n)$ is related to the diagonal elements $[{\bf \Sigma}_{\sigma}(i\omega_n)]_{ii}$ of the matrix self-energy and the diagonal elements $[{\bf G}_{\sigma}(i\omega_n)]_{ii} $ of  the one-particle matrix Green function through the local Dyson equation
\begin{equation}
{\cal G}_{i\sigma}(i\omega_n)=\left[[{\bf G}_{\sigma}^{-1} (i\omega_n)]_{ii}+ [{\bf \Sigma}_{\sigma}(i\omega_n)]_{ii}\right]^{-1}.
\label{Dyson_local}
\end{equation}
These equations allow one to determine all diagonal elements of the matrix Green function at each lattice site.
In real space the Dyson equation has the form
\begin{equation}
{\bf G}^{-1}_{\sigma}(i\omega_n) = (i\omega_n+\mu){\bf 1} - {\bf H}_0 - {\bf V}_{\sigma} - {\bf \Sigma}_{\sigma}(i\omega_n).
\label{Dyson}
\end{equation}
Here we used a matrix (bold face) notation where the matrix element $[{\bf G}_{\sigma}(i\omega_n)]_{ij}$  is the one-particle Green function of the interacting electrons propagating from site $i$ to $j$, and the matrix element $[{\bf \Sigma}_{\sigma}(i\omega_n)]_{ij}$ is the corresponding self-energy which takes all interaction effects in the system into account and which is diagonal within the R-DMFT, i.e. $[{\bf \Sigma}_{\sigma}(i\omega_n)]_{ij}=\delta_{ij} [{\bf \Sigma}_{\sigma}(i\omega_n)]_{ii}$. The other matrix elements are $[{\bf 1}]_{ij}=\delta_{ij}$, $[{\bf H}_0]_{ij}=t_{ij}$, and
$ [{\bf V}_{\sigma}]_{ij}= \delta_{ij} V_{i\sigma}$.
The superscript ${-1}$ at a bold face symbol means matrix inversion.

\section{Scattering formulation of the R-DMFT solution}

For a homogeneous system, i.e. when $V_{i\sigma}=0$, the R-DMFT leads to  the same self-energy ${\bf \Sigma}_{\sigma}(i\omega_n)_{ij}=\delta_{ij}  \Sigma_{\sigma}^0(i\omega_n)$ on each lattice site.
The scalar self-energy $\Sigma_{\sigma}^0(i\omega_n)$  describes all interaction effects in the homogeneous system.
In an {\em in}homogeneous system we can split the matrix self-energy as follows:
\begin{equation}
[{\bf \Sigma}_{\sigma}(i\omega_n)]_{ij}=\delta_{ij}  \Sigma_{\sigma}^0(i\omega_n) + \delta_{ij} [{\bf \Delta \Sigma}_{\sigma}(i\omega_n)]_{ii},
\label{selfenergy}
\end{equation}
where within the R-DMFT $ {\bf \Delta \Sigma}_{\sigma}(i\omega_n) $ is a diagonal matrix which takes into account the  interaction effects in the presence of  the external potential $V_{i\sigma}\neq 0$.\cite{Ziegler96,Lederer08}
This site-dependent part of the self-energy vanishes when $V_{i\sigma} =0$.
With this separation we rewrite the Dyson equation (\ref{Dyson}) in the form
\begin{equation}
{\bf G}^{-1}_{\sigma}(i\omega_n) = {\bf G}^{-1}_{{\rm hom}\;\sigma}(i\omega_n) - {\bf \tilde V}_{\sigma}(i\omega_n),
\label{Dyson_free}
\end{equation}
where the one-particle Green function in the homogeneous, interacting system is defined as
\begin{equation}
{\bf G}^{-1}_{{\rm hom}\;\sigma}(i\omega_n) = [i\omega_n+\mu -\Sigma_{\sigma}^0(i\omega_n)]{\bf 1} - {\bf H}_0.
\label{Dysonhom}
\end{equation}
Here
\begin{equation}
{\bf \tilde V}_{\sigma}(i\omega_n) = {\bf V}_{\sigma} + {\bf \Delta \Sigma}_{\sigma}(i\omega_n)
\label{potential}
\end{equation}
is a dynamical potential acting on the particles, which is  due to the presence of the external perturbation $V_{i\sigma}$ and the electron interaction $U$.
The first term on the r.h.s of (\ref{potential}), ${\bf V}_{\sigma}$, is static and may be interpreted as a scattering amplitude arising from the external perturbation.
By contrast, the second term is due to the screening processes in the interacting system.
Since the dynamical part of the self-energy
vanishes for $\omega_n\rightarrow \pm \infty$  only the screened static part of the potential influences the high energy states, i.e.
\begin{equation}
{\bf \tilde V}_{\sigma}(i\omega_n \rightarrow \pm \infty) = {\bf V}_{\sigma} + U {\bf \Delta N},
\end{equation}
where $[{\bf \Delta N}]_{ij}=\delta_{ij} (\bar{n}_{i\sigma}-\bar{n}_{{\rm hom}\; \sigma}) \equiv \delta_{ij}\Delta n_{i\sigma}$ is a diagonal matrix describing the deviation of local occupations $\bar{n}_{i\sigma}$ with respect to the homogeneous occupation $\bar{n}_{{\rm hom}\; \sigma}$.

In view  of the formal similarity between the relation (\ref{Dyson_free}) and the corresponding resolvent equation in the theory of scattering problems, Eq.~(\ref{Dyson_free}) can be solved by introducing the dynamical T-matrix\cite{Ziegler96,Lederer08,Mukherjee15,Doniach}
\begin{eqnarray}
{\bf T}_{\sigma}(i\omega_n) = [{\bf 1} - {\bf \tilde V}_{\sigma}(i\omega_n) {\bf G}_{{\rm hom}\;\sigma}(i\omega_n)]^{-1} {\bf \tilde V}_{\sigma}(i\omega_n)  \nonumber\\
= {\bf \tilde V}_{\sigma}(i\omega_n) [{\bf 1} - {\bf G}_{{\rm hom} \;\sigma}(i\omega_n) {\bf \tilde V}_{\sigma}(i\omega_n) ]^{-1}.
\label{solution-t}
\end{eqnarray}
We note that for $\omega_n\rightarrow \pm \infty$ the T-matrix reduces to the matrix of the screened external potential, i.e., ${\bf T}_{\sigma}(i\omega_n) \rightarrow {\bf V}_{\sigma}+ U {\bf \Delta N}$ because ${\bf G}_{{\rm hom}\;\sigma }(i\omega_n)\sim 1/i\omega_n$ in this limit.
The solution of (\ref{Dyson_free}) then takes the form
\begin{eqnarray}
{\bf G}_{\sigma}(i\omega_n) = {\bf G}_{{\rm hom} \;\sigma}(i\omega_n) \nonumber \\
 + {\bf G}_{{\rm hom} \;\sigma}(i\omega_n) {\bf T}_{\sigma}(i\omega_n) {\bf G}_{{\rm hom}\;\sigma }(i\omega_n).
\label{solution}
\end{eqnarray}
In the case of non-interacting electrons ($U=0$) and an external potential which is either local or has, at most, a finite range, the T-matrix (\ref{solution-t}) has only few non-vanishing elements.
For example, for a point-like impurity potential ${\bf V}_{ij}= \delta_{ij} \delta_{ii_0}V_0$ located at site $i_0$ there is only a single non-vanishing matrix element.
Then  the full Green function in Eq.~(\ref{solution}) may be obtained simply by algebraic multiplication of finite matrices.
In the interacting case ($U\neq 0$) the problem is more difficult since the dynamical potential ${\bf \tilde V}_{\sigma}(i\omega_n)$ is long-ranged even if the external potential $ {\bf V}_{\sigma}$ is not.

\section{Density of states and scattering phase shifts}

The local density of states (LDOS) provides direct information about the local occupation of particular lattice sites. It is given by
\begin{equation}
\rho_{i\sigma}(\omega) = - \frac{1}{\pi} {\rm Im}\;  [{\bf G}_{\sigma}(\omega + i\eta)]_{ii},
\end{equation}
where the analytical continuation from Matsubara to real frequencies was performed, i.e., $i\omega_n \rightarrow \omega +i\eta$ with $\eta \rightarrow 0^+$, and $\rm Im$ refers to the imaginary part.
Hence, the total density of state (DOS) is
\begin{equation}
\rho(\omega)= \sum_{i\sigma} \rho_{i\sigma}(\omega) = -\frac{1}{\pi} {\rm Im}\; {\rm Tr} {\bf G}_{\sigma}(\omega + i\eta),
\end{equation}
where the trace ($\rm Tr$) denotes summation over the diagonal elements, i.e., over all lattice sites and spin indices.
In particular, we are interested in the changes of the DOS and the particle number in the system due to the external potential.
With
\begin{eqnarray}
\Delta \rho_{i\sigma} (\omega) = \rho_{i\sigma}(\omega) - \rho_{{\rm hom} \;\sigma}(\omega)
\end{eqnarray}
we compute the change of the DOS
\begin{align}
&\Delta \rho (\omega) = \sum_{i\sigma} \Delta \rho_{i\sigma} (\omega) \nonumber \\
&= - \frac{1}{\pi} {\rm Im} \; {\rm Tr} \left[ {\bf G}_{\sigma}(\omega + i\eta) - {\bf G}_{{\rm hom} \;\sigma}(\omega + i\eta)\right].
\label{changedos}
\end{align}

Using the identity for the matrix Green function ${\bf G}_{\sigma}(z)$ of complex argument $z$ (see Appendix A)
\begin{equation}
{\rm Tr} {\bf G}_{\sigma}(z) = \frac{d}{d z}  {\rm Tr} \ln {\bf G}_{\sigma}^{-1}(z)+ {\rm Tr}\left[ {\bf G}_{\sigma}(z) \frac{d {\bf \Sigma}_{\sigma}(z)}{d z} \right]
\label{trace_G}
\end{equation}
together with Eqs.~(\ref{selfenergy}), (\ref{potential}) and the Dyson equations ({\ref{Dyson})~and~(\ref{Dysonhom}), we obtain (see Appendix B)
\begin{align}
&{\rm Tr} \left[ {\bf G}_{\sigma}(z) - {\bf G}_{{\rm hom} \;\sigma}(z) \right] = \nonumber \\
&\frac{d}{dz} {\rm Tr} \left[  \ln [{\bf 1} - {\bf \tilde V}_{\sigma}(z) {\bf G}_{{\rm hom} \;\sigma}(z)] \right]   \nonumber \\
&+ {\rm Tr}\left[ {\bf G}_{\sigma}(z) \frac{d {\bf \Sigma}_{\sigma}(z)}{d z} \right]
-{\rm Tr}\left[ {\bf G}_{{\rm hom}\;\sigma}(z) \frac{d \Sigma^0_{\sigma}(z)}{d z} \right] .
\label{trace_(G-Ghom)}
\end{align}
In terms of the T-matrix (\ref{solution-t}) this reads
\begin{align}
&{\rm Tr} \left[ {\bf G}_{\sigma}(z) - {\bf G}_{{\rm hom} \;\sigma}(z) \right] \nonumber \\
&= - \frac{d}{dz} {\rm Tr} \left[  \ln [
{\bf \tilde V}_{\sigma}^{-1}(z) {\bf T}_{\sigma}(z)] \right]   \nonumber \\
&+ {\rm Tr}\left[ {\bf G}_{\sigma}(z) \frac{d {\bf \Sigma}_{\sigma}(z)}{d z} \right]
-{\rm Tr}\left[ {\bf G}_{{\rm hom}\;\sigma}(z) \frac{d \Sigma^0_{\sigma}(z)}{d z} \right] .
\end{align}
Next we define the  matrix of  scattering phase shifts as
\begin{eqnarray}
{\boldsymbol \phi}_{\sigma}(z)= {\rm Im} \ln [{\bf \tilde V}_{\sigma}^{-1}(z) {\bf T}_{\sigma}(z)]\nonumber \\=
{\rm Arg}[ {\bf \tilde V}_{\sigma}^{-1}(z) {\bf T}_{\sigma}(z)],
\label{matrix-shifts}
\end{eqnarray}
where $\rm Arg$ refers to the argument of the complex quantity.\cite{comment,comment-bis}

The change of the DOS then takes the familiar form
\begin{align}
&\Delta \rho (\omega) =   \frac{1}{\pi} {\rm Tr} \left[ \frac{d {\boldsymbol \phi}_{\sigma} (\omega+ i\eta) }{d \omega}
 \nonumber \right.\\ -
 &{\rm Im} \; \left( {\bf G}_{\sigma}(\omega+ i\eta) \frac{d {\bf \Sigma}_{\sigma}(\omega+ i\eta)}{d \omega}\right) \nonumber \\
&\left. +{\rm Im}\;\left(  {\bf G}_{{\rm hom} \;\sigma}(\omega+ i\eta) \frac{d  \Sigma^0_{\sigma}(\omega+ i\eta)}{d \omega} \right)
\right].
\label{delta-rho}
\end{align}
Since ${\bf T}_{\sigma}(\omega)\rightarrow {\bf V}_{\sigma} + U{\bf \Delta N}$ for $\omega \rightarrow \pm \infty$ the matrix of scattering phase shifts obeys the limit ${\boldsymbol \phi}_{\sigma}(\omega) \rightarrow 0$ and likewise $\Delta \rho (\omega) \rightarrow 0$.\cite{boundstates}
We note that in the non-interacting case ($U=0$) the standard expression  $ \Delta \rho (\omega) =  -\frac{1}{\pi} {\rm Tr} [\frac{d {\boldsymbol \phi}_{\sigma} (\omega) }{d \omega}]$ is recovered.

Since ${\bf T}_{\sigma}(\omega)\rightarrow {\bf V}_{\sigma} + U{\bf \Delta N}$ for $\omega \rightarrow \pm \infty$ the matrix of scattering phase shifts obeys the limit ${\boldsymbol \phi}_{\sigma}(\omega) \rightarrow 0$ and likewise $\Delta \rho (\omega) \rightarrow 0$.\cite{boundstates}
We note that in the non-interacting case ($U=0$) the standard expression  $ \Delta \rho (\omega) =  -\frac{1}{\pi} {\rm Tr} [\frac{d {\boldsymbol \phi}_{\sigma} (\omega) }{d \omega}]$ is recovered.

\section{Friedel sum rule}

The  change of the local occupation is directly determined from the LDOS as
\begin{eqnarray}
\Delta n_{i\sigma} = \int_{-\infty}^{+\infty} d \omega f(\omega)  \Delta \rho_{i\sigma}(\omega),
\label{lococc}
\end{eqnarray}
where $f(\omega)=1/(\exp(\beta \omega)+1)$ is the Fermi-Dirac function.
At $T=0$ Eq.~(\ref{lococc})  reduces to
\begin{equation}
\Delta n_{i\sigma} = \int_{-\infty}^{0} d \omega \Delta \rho_{i\sigma}(\omega).
\end{equation}
The integral extends to zero because the chemical potential is already included in the Green function.

The total change of the site occupations provides the screening charge $Z_{\rm sc}$ as
\begin{eqnarray}
Z_{\rm sc} = \sum_{i\sigma} \Delta n_{i\sigma} = \int_{-\infty}^{+\infty} d \omega f(\omega) \Delta \rho(\omega),
\end{eqnarray}
where $\Delta\rho(\omega)$ was defined in the first line of Eq.~(\ref{changedos}).

Finally, using (\ref{delta-rho})  the screening charge is found as
\begin{align}
&Z_{\rm sc}
=  \frac{1}{\pi}  \int_{-\infty}^{+\infty} d \omega f(\omega) \left(
\frac{d}{d\omega} {\rm Tr} \left[  {\boldsymbol \phi}_{\sigma}(\omega)  \right]  \right.  \nonumber \\
&- \;{\rm Im} {\rm Tr} \left[ {\bf G}_{\sigma}(\omega+ i\eta) \frac{d {\bf \Sigma}_{\sigma}(\omega+ i\eta)}{d \omega} \right] \nonumber \\
&+\left. {\rm Im}  {\rm Tr} \left[ {\bf G}_{{\rm hom}\;\sigma}(\omega+ i\eta) \frac{d \Sigma^0_{\sigma}(\omega+ i\eta)}{d \omega} \right]  \right)  .
\end{align}
At $T=0$ it can be expressed by the scattering phase shift matrix as
\begin{equation}
Z_{\rm sc} =  \frac{1}{\pi} {\rm Tr} \int_{-\infty}^{0} d \omega \frac{d {\boldsymbol \phi}_{\sigma} (\omega) }{d \omega} ,
\label{last}
\end{equation}
where we used the exact expression due to Luttinger\cite{Luttinger_bis60,Luttinger60}
\begin{equation}
 \frac{1}{\pi} {\rm Tr} \left[ \int_{-\infty}^{0} d \omega {\rm Im} \; {\bf G}_{\sigma}(\omega) \frac{d {\bf \Sigma}_{\sigma}(\omega)}{d \omega}\right] =0
\label{Luttinger_integral}
\end{equation}
based on the conservation of the total number of particles derived in Appendix C.\cite{Oshikawa00,Seki17}
The same relation holds for ${\bf G}_{{\rm hom} \;\sigma}(\omega)$.
By integration of (\ref{last}) we obtain the standard form of the Friedel sum rule
\begin{equation}
Z_{\rm sc} = \frac{1}{\pi} {\rm Tr}\; {\boldsymbol \phi}_{\sigma} (0),
\label{friedel-zero}
\end{equation}
where ${\boldsymbol \phi}_{\sigma} (0)$ is the scattering phase shift matrix at  the Fermi energy $\epsilon_F$.\cite{boundstates}
The Friedel sum rule is seen to be an algebraic  relation between the screening charge and the phase shifts even in interacting lattice fermions.\cite{Langer61}
To illustrate this result we discuss in the Appendix D an analytic  solution for a single impurity within a homogeneous self-energy approximation.

\section{Summary and outlook}
We reformulated the real-space dynamical mean-field theory (R-DMFT) in the framework of scattering theory by introducing a T-matrix  and a  dynamical potential. The latter describes scattering processes due to an external inhomogeneity $V_{i\sigma}$ and the interaction $U$. This allowed us to rewrite the Dyson equation (\ref{Dyson}) in a form which is familiar from  scattering theory of quantum mechanics. By defining the  matrix of scattering phase shifts we derived the Friedel sum rule (which, in principle, holds for general many-body systems \cite{Langer61}) for the case of interacting lattice fermions within R-DMFT.

The T-matrix formulation can be used to develop new numerical approaches to solve or approximate the R-DMFT equations. For example, when the self-energy corrections $[{\bf \Delta \Sigma}_{\sigma}(i\omega_n)]_{ii}$ to the dynamical potential (\ref{potential}) are neglected, one obtains a homogeneous self-energy approximation for R-DMFT.\cite{Chatterjee17} Furthermore, the exact Friedel sum rule for the lattice problem can be employed to assess the quality of  numerical or approximate solutions of R-DMFT.

Finally we note that the algebraic results presented in Sections IV, V, and VI are in fact valid for an \emph{arbitrary} exact self-energy $\Sigma_{ij\sigma}(\omega)$ when the latter is written as a sum of homogeneous and inhomogeneous parts, i.e.,  $\Sigma_{ij\sigma}(\omega)=\Sigma^0_{\sigma}(\omega)\delta_{ij} + \Delta \Sigma_{ij\sigma}(\omega)$.
Thereby the Friedel sum rule is found to be valid for an arbitrary system of correlated lattice fermions in the Fermi liquid state at $T=0$.
We should like to stress that our formulation is purely in position space and does not refer to momentum space at all.

The R-DMFT framework discussed here provides an efficient tool for determining an approximate self-energy when the Friedel sum rule holds.

\begin{acknowledgments}

We thank  A. Weh and J.~Skolimowski for useful comments and discussions.
Support by the Deutsche Forschungsgemeinschaft through TRR 80 is gratefully acknowledged. BC also acknowledges discussions with J. Kolorenc as well as financial support from the Czech Academy of Sciences.

\end{acknowledgments}

\appendix

\section{Derivation of Eq.~(\ref{trace_G})}

To derive Eq.~(\ref{trace_G}) we use the following Jacobi identity \cite{Jacobi} for an invertible  finite matrix function ${\bf B}(z)$:
\begin{equation}
\frac{d}{dz}  {\rm Det}{\bf B} (z)= {\rm Det} {\bf B}(z) {\rm Tr} \left[ {\bf B}^{-1}(z) \frac{{\bf B} (z)}{dz } \right],
\end{equation}
where ${\rm Det}$ denotes the determinant.
Then, together with
\begin{equation}
{\rm Tr }\ln {\bf B}(z)=\ln {\rm Det}{\bf B}(z),
\label{trace-log}
\end{equation}
we compute
\begin{align}
\frac{d}{dz}&{\rm Tr} \ln {\bf G}_{\sigma}^{-1}(z)\nonumber\\
&= \frac{1}{ {\rm Det}{\bf G }_{\sigma}^{-1} (z)} \frac{d}{dz}  {\rm Det}{\bf G}_{\sigma}^{-1} (z) \nonumber \\
&={\rm Tr}\left[  {\bf G}_{\sigma}(z) \frac{d}{dz} {\bf G}_{\sigma}^{-1}(z) \right] \nonumber \\
&={\rm Tr}\left[  {\bf G}_{\sigma}(z) \frac{d}{dz}
[(z+\mu){\bf 1} -{\bf H}_0-{\bf V}_{\sigma} -{\bf \Sigma}_{\sigma}(z)  ]
 \right] \nonumber \\
&= {\rm Tr}\left[  {\bf G}_{\sigma}(z)
[{\bf 1} -\frac{d}{dz} {\bf \Sigma}_{\sigma}(z)  ]
 \right] .
\end{align}
From the linearity of the trace, ${\rm Tr}({\bf A}+{\bf B})= {\rm Tr}{\bf A}+ {\rm Tr}{\bf B}$, we obtain eq. (\ref{trace_G}).
A corresponding identity holds for ${\bf G}_{{\rm hom}\;\sigma}(z)$.

\section{Derivation of Eq.~(\ref{trace_(G-Ghom)})}

With Eq.~(\ref{trace_G}) we obtain
\begin{multline}
{\rm Tr} [{\bf G}_{\sigma}(z) - {\bf G}_{{\rm hom} \;\sigma}(z)] = \\
\frac{d}{dz} \left[ {\rm Tr} \ln {\bf G}_{\sigma}^{-1} (z) - {\rm Tr} \ln {\bf G}_{{\rm hom} \; \sigma}^{-1} (z)
\right] \\
+ {\rm Tr}\left[ {\bf G}_{\sigma}(z) \frac{d {\bf \Sigma}_{\sigma}(z)}{d z} \right]
-{\rm Tr}\left[ {\bf G}_{{\rm hom}\;\sigma}(z) \frac{d \Sigma^0_{\sigma}(z)}{d z} \right] .
\end{multline}
Using the identity Eq.~(\ref{trace-log}) together with ${\rm Det} {\bf A}  {\rm Det}{\bf B}={\rm Det}{(\bf AB)}$
we find
\begin{align}
{\rm Tr} \ln {\bf G}_{\sigma}^{-1} (z) &- {\rm Tr} \ln {\bf G}_{{\rm hom} \; \sigma}^{-1} (z) \nonumber \\
&=\ln {\rm Det}{\bf G}_{\sigma}^{-1} (z) - \ln {\rm Det} {\bf G}_{{\rm hom} \; \sigma}^{-1} (z) \nonumber \\
&= \ln \left[{\rm Det}{\bf G}_{\sigma}^{-1} (z)\: {\rm Det} {\bf G}_{{\rm hom} \; \sigma} (z)\right] \nonumber \\
&=  \ln \Big[ {\rm Det} \left[ {\bf G}_{\sigma}^{-1} (z) {\bf G}_{{\rm hom} \; \sigma} (z)\right]  \Big]
 \nonumber \\
&= {\rm Tr}  \ln  \left[{\bf G}_{\sigma}^{-1} (z) {\bf G}_{{\rm hom} \; \sigma} (z)\right].
\label{trace-log-2}
\end{align}
Finally, using Eqs.~({\ref{Dyson})~and~(\ref{Dysonhom}) we rewrite the last line of Eq.~(\ref{trace-log-2}) as
\begin{align}
{\rm Tr}  \ln & \big[ {\bf G}_{\sigma}^{-1} (z) {\bf G}_{{\rm hom} \; \sigma} (z) \big] \nonumber \\
&={\rm Tr} \ln \Big[
\big[
(z+\mu){\bf 1} -{\bf H}_0-{\bf V}_{\sigma} -{\bf \Sigma}_{\sigma}(z)
\big]
\nonumber \\
&~~~~~~~~~~~~
\cdot
\big[
(z+\mu -\Sigma_{\sigma}^0(z)){\bf 1} - {\bf H}_0
\big]^{-1}
\Big] \nonumber \\
&= {\rm Tr}   \ln \Big[ {\bf 1} - [{\bf V}_{\sigma}+ {\bf \Delta \Sigma}_{\sigma}(z) ]
\nonumber \\
&~~~~~~~~~~~~
\cdot
[(z+\mu -\Sigma_{\sigma}^0(z)){\bf 1} - {\bf H}_0 ]^{-1}
\Big] \nonumber \\
&={\rm Tr}  \ln [{\bf 1} - {\bf \tilde V}_{\sigma}(z) {\bf G}_{{\rm hom} \;\sigma}(z)],
\end{align}
where in the last two steps we used Eqs.~(\ref{selfenergy}) and (\ref{potential}), respectively.
Combining (B2) and (B3) with (B1) we obtain Eq.~(\ref{trace_(G-Ghom)}).

\section{Derivation of Eq.~(\ref{Luttinger_integral})}

To derive Eq.~(\ref{Luttinger_integral}) we consider the following sum
\begin{align}
I(T) =- \frac{1}{\beta} \sum_{\omega_n} {\rm Tr} \left[{\bf G}_{\sigma}(i\omega_n) \right. \nonumber \\
\left. \cdot \frac{ {\bf \Sigma}_{\sigma} (i\omega_n+i \delta)  - {\bf \Sigma}_{\sigma} (i\omega_n) }{  \delta }  \right] e^{i\omega_n \eta},
\label{Sum}
\end{align}
where $\delta \equiv \pi/\beta \rightarrow 0$ when $T\rightarrow 0$ and $\eta \rightarrow 0^+$.
Using the relation\cite{Bruus}
\begin{equation}
-\frac{1}{\beta} \sum_{\omega_n} g(\omega_n)e^{i\omega_n\eta}= \frac{1}{\pi} \int_{-\infty}^{\infty} d \omega f(\omega) {\rm Im} g(\omega) e^{i\omega \eta},
\end{equation}
where $g(z)$ is a general function which is  non-analytic only on the real axis (in general it has a branch cut on the real axis) and $f(\omega)$ is the Fermi-Dirac function, we find
\begin{align}
I(T) = \frac{1}{\pi} {\rm Tr}  \int_{-\infty}^{\infty} d \omega f(\omega){\rm Im}  \left[{\bf G}_{\sigma}(\omega) \right. \nonumber \\
\left. \cdot \frac{ {\bf \Sigma}_{\sigma} (\omega+\delta)  - {\bf \Sigma}_{\sigma} (\omega) }{   \delta }  \right] e^{i\omega_n \eta}.
\end{align}
At zero temperature we find
 \begin{equation}
I(T=0) = \frac{1}{\pi} {\rm Tr}  \int_{-\infty}^{0} d \omega {\rm Im}  \left[{\bf G}_{\sigma}(\omega) \frac{ d {\bf \Sigma}_{\sigma} (\omega)  }{  d  \omega }  \right] ,
\end{equation}
which is the integral in (\ref{Luttinger_integral}).

Furthermore, at $T=0$ and after performing an integration by parts the sum (\ref{Sum})  takes the form\cite{integral}
\begin{equation}
I(T=0) =  \int_{-\infty}^{\infty}  \frac{d(i\omega) }{2\pi} {\rm Tr} \left[{\bf \Sigma}_{\sigma}(i\omega) \frac{d {\bf G}_{\sigma}(i\omega)}{d (i\omega)} \right].
\label{luttinger_int}
\end{equation}
Since ${\bf G}_{\sigma}(i\omega_n)\sim 1/i\omega_n$ at large Matsubara frequencies we see that the boundary terms
$\lim_{\omega\rightarrow \pm \infty} {\bf G}_{\sigma}(i\omega) {\bf \Sigma}_{\sigma}(i\omega) =0$. It remains to be shown that $I(T=0) =0$.\cite{Ward-functional}

In a normal-state system where the number of particles is conserved, the particle operator $\hat{N}=\sum_{i\sigma} \hat{n}_{i\sigma}$  commutes not only with the total Hamiltonian but also with its parts, i.e.,
\begin{equation}
0=[\hat{H},\hat{N}]= [\hat{H}_0,\hat{N}] + [\hat{H}_{\rm int},\hat{N}] .
\label{commutators}
\end{equation}
We will now show that for systems where (\ref{commutators}) holds $I(T=0) = 0$ in (\ref{luttinger_int}).
Following Fabrizio\cite{Fabrizio}  we consider a unitary symmetry transformation $\hat{U}(\delta)= e^{i\delta \hat{N} \tau}$, with $\delta=\pi/\beta$ as before.  It is easy to check that $\hat{H}_{\rm int}(\tau)=e^{\hat{H}_0\tau}\hat{H}_{\rm int}e^{-\hat{H}_0\tau}$ is invariant under this symmetry, i.e., $ \hat{U}(\delta)\hat{H}_{\rm int}(\tau) \hat{U}^{\dagger} (\delta)=\hat{H}_{\rm int}(\tau)$.
This implies that the Ward functional $\Phi[{\bf G}_{\sigma}]$, expressed as an infinite sum of time-ordered integrals of $\langle   \hat{H}_{\rm int}(\tau_1)....\hat{H}_{\rm int}(\tau_n)) \rangle  $, is invariant under this symmetry operation as well.

From the property $\hat{U}(\delta)\hat{a}_{i\sigma}(\tau)\hat{U}^{\dagger}(\delta)=e^{-i\delta \tau}\hat{a}_{i\sigma}(\tau)$ it follows that
$G_{ij\sigma}(\tau)$ transforms under the symmetry operation $\hat{U}(\delta)$ into $e^{-i\delta \tau }G_{ij\sigma}(\tau)$ and the corresponding Fourier component $G_{ij\sigma}(i\omega_n)$ into $G_{ij\sigma}(i\omega_n+i\delta)$. Then we can compute the difference of the Ward functionals as
\begin{align}
\frac{\Delta \Phi[{\bf G}_{\sigma}]}{\delta} = \frac{\Phi[{\bf G}_{\sigma}(i\omega_n + i\delta )] -  \Phi[{\bf G}_{\sigma}(i\omega_n)] }{\delta} \nonumber \\
=  \frac{1}{\beta} \sum_{\omega_n} {\rm Tr } \left[   {\bf \Sigma}_{\sigma}(i\omega_n)  \frac{{\bf G}_{\sigma}(i\omega_n + i\delta) - {\bf G}_{\sigma}(i\omega_n)] }{\delta }   \right] \nonumber \\
\xrightarrow[T\rightarrow 0]{ }
 \int_{-\infty}^{\infty}  \frac{d(i\omega)}{2\pi} {\rm Tr} \left[ {\bf \Sigma}_{\sigma} (i\omega) \frac{d{\bf G}_{\sigma}(i\omega)}{d(i\omega)} \right] =0,
\label{luttinger_end}
\end{align}
where the last equality follows from the fact  that $\Delta \Phi[{\bf G}_{\sigma}]=0$ under the symmetry operation $\hat{U}(\delta)$ (see the discussion below (\ref{commutators})).
Comparing (\ref{luttinger_end}) with (\ref{luttinger_int})  we conclude that $I(T=0)=0$.

The reasoning presented above is based on a conservation law.\cite{Fabrizio} By contrast, following the approach by Luttinger\cite{Luttinger_bis60}, Abrikosov, Gorkov, and Dzyaloshinski\cite{AGD} argued from the point of view of the diagrammatic structure of the Ward functional and its invariance upon a shift of the frequency of the Green functions.\cite{footnote} We see that both points of view are equivalent.


\section{Analytic solution for a single-site impurity within a homogeneous self-energy approximation}

As an illustration we consider the scattering from an  impurity located at site $i=1$, where  the interaction is treated  by assuming a homogeneous self-energy, i.e., where the additional term ${\bf \Delta \Sigma}_{\sigma}(i\omega_n)$ in Eq.~(\ref{selfenergy}) is neglected ("homogeneous self-energy approximation" (HSEA)).\cite{Chatterjee17}
Then the dynamical potential (\ref{potential})  is a matrix with only one non-vanishing element
\begin{equation}
{\bf \tilde V}_{\sigma}(\omega)=\left(
\begin{array}{cccc}
V&0&0&...\\
0&0&0&...\\
...&...&...&...
\end{array}
\right).
\end{equation}
In this case the T-matrix is simply given by
\begin{equation}
{\bf T}_{\sigma}(\omega)=\frac{1}{1-V [{\bf G}_{\rm hom \;\sigma}(\omega)]_{11}}\left(
\begin{array}{cccc}
V&0&0&...\\
0&0&0&...\\
...&...&...&...
\end{array}
\right),
\end{equation}
i.e., also has only one non-vanishing element.
Within the HSEA for the R-DMFT all correlation effects are taken into account by a single scalar self-energy determined by the regular DMFT and $  {\bf G}_{\rm hom \;\sigma}(\omega) = {\bf G}_{\sigma}^{0}(\omega+\mu-\Sigma_{\sigma}^0(\omega))$, where ${\bf G}_{\sigma}^{0}(\omega)$ is the non-interacting Green function.
Now the matrix of scattering phase shifts (\ref{matrix-shifts}) consists of only  one diagonal element
\begin{equation}
[{\boldsymbol \phi}_{\sigma}(\omega)]_{11}= -{\rm arctan} \left(\frac{V {\rm Im}  [{\bf G}_{\rm hom \;\sigma}(\omega)]_{11} }{1- V {\rm Re} [{\bf G}_{\rm hom \;\sigma}(\omega)]_{11} } \right),
\end{equation}
where $\rm Re$ referes to the real part.
The Friedel sum rule (\ref{friedel-zero}) then reads
\begin{align}
Z_{\rm sc}=-\frac{1}{\pi} {\rm arctan} \left(\frac{V {\rm Im}  [{\bf G}_{\rm hom \;\sigma}(0)]_{11} }{1- V {\rm Re} [{\bf G}_{\rm hom \;\sigma}(0)]_{11} } \right) \nonumber \\
\nonumber \\
= -\frac{1}{\pi} {\rm arctan} \left(\frac{V {\rm Im}  G_{\sigma}^0(\mu-\Sigma_{\sigma}^0(0) ) }{1- V {\rm Re}  G_{\sigma}^0(\mu-\Sigma_{\sigma}^0(0) )} \right) .
\label{doniach-general}
\end{align}
For non-interacting systems with $\Sigma_{\sigma}^0=0$ the second equation in (\ref{doniach-general}) reduces to the well-known result found in the literature, see e.g.~Ref.~[\onlinecite{Doniach}].
For interacting systems described by the R-DMFT within the HSEA eq. (\ref{doniach-general}) is an explicit analytical results for the Friedel sum rule, which holds even in the presence of correlations.


\begin{thebibliography}{}
\bibitem{comment-bis} In non-interacting systems this definition of the scattering phase shifts is equivalent to the standard definition via the scattering S-matrix. Therefore, these quantities  would have the standard interpretation of phase shifts of the asymptotic wave functions. In interacting systems such an interpretation is no longer valid because the asymptotic one-particle scattering states cannot be defined in general. Nevertheless, the  matrix of scattering phase shifts introduced here turns out to be a useful object in analyzing correlated many-body systems as will be seen below.
\bibitem{integral} We use the relation $(1/ \beta) \sum_{\omega_n} = (1/ 2\pi) \sum_{\omega_n} \Delta \omega_n$, where $\Delta \omega_n= \omega_{n+1}-\omega_n=2\pi/\beta$, and in the limit $\beta\rightarrow \infty$ ($T\rightarrow 0$) we replace the sum by the integral $(1/ \beta) \sum_{\omega_n=-\infty}^{\infty} \rightarrow _{T\rightarrow 0} \int_{-\infty}^{\infty}  d\omega/2\pi$.
\bibitem{Ward-functional} We note that the standard arguments leading to $I(T=0)=0$  assume the existence of the Ward functional  $ \Phi [{\bf G}_{\sigma}]$  as a converged sum of all skeleton diagrams, such that ${\bf \Sigma}_{\sigma} (i\omega_n) = \delta \Phi[{\bf G}_{\sigma}] / \delta {\bf G}_{\sigma}(i\omega_n)$. It then follows that $ \Phi[{\bf G}_{\sigma}]= (1/\beta) \sum_{\omega_n} {\rm Tr} [{\bf \Sigma}_{\sigma }(i\omega_n) {\bf G}_{\sigma}(i\omega_n)]$.
\bibitem{footnote} This is because at $T=0$ the integration limits are $\pm \infty$, and at each vertex there is a $\delta$-function with the same number of frequencies for incoming and outgoing  lines.
\end{thebibliography}
\end{document}